Symmetry and financial Markets

Jørgen Vitting Andersen* and Andrzej Nowak**

July 2020

* CNRS, Centre d'Economie de la Sorbonne, Université Paris 1 Pantheon-Sorbonne, Maison des Sciences Economiques,106-112 Boulevard de l'Hôpital, 75647 Paris Cedex 13, France. Email: jorgen-vitting.andersen@univ-paris1.fr

** Department of Psychology, Warsaw University, Warsaw, Poland

**Abstract**

It is hard to overstate the importance that the concept of symmetry has had in every field of physics, a fact alluded to by the Nobel Prize winner P.W. Anderson, who once wrote that "physics is the study of symmetry". Whereas the idea of symmetry is widely used in science in general, very few (if not almost no) applications has found its way into the field of finance. Still, the phenomenon appears relevant in terms of for example the symmetry of strategies that can happen in the decision making to buy or sell financial shares. Game theory is therefore one obvious avenue where to look for symmetry, but as will be shown, also technical analysis and long term economic growth could be phenomena which show the hallmark of a symmetry.

1. **Introduction**

Since this article is meant for an interdisciplinary audience (for a longer essay on interdisciplinary approaches to finance, see also [1,2]), it is probably in order to begin the article by illustrating the concept of symmetry. Take a look at figure 1a where you can imagine supporting a pen held upright on a table with a finger. In the case illustrated there is a 360 degrees' rotational symmetry, since, after lifting your finger, the pen can fall in any direction.

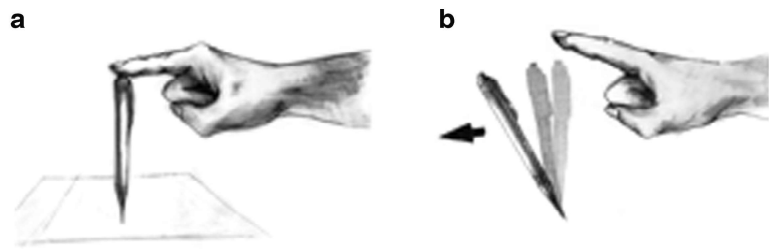

**Figure 1: Symmetry, and the breaking of symmetry.** *A pen falls and a symmetry is broken. Fig. 1a illustrates a symmetric state of a pen, since in this state there is no preferred orientation, and the pen can fall in any direction. As soon as you lift the finger, a tiny fluctuation will force the pen to choose a given direction, and the symmetric is broken.*

After you slowly lift your finger, the pen suddenly falls in an arbitrary direction as depicted in figure 1b. Once the pen begins its fall, a random direction has been chosen/fixed: you don't see the pen suddenly change and flip back in another direction. One then says that the 360 degrees' rotational symmetry is "broken".

For physicists the idea of symmetry is familiar, used in all sub-disciplines of the field. For example, properties of particles originate from the symmetries of laws of physics [3]. The concept of symmetry also plays a fundamental role in other sciences for example mathematics, biology, chemistry, and neuroscience. In finance, however, the concept of symmetry has not yet been adopted.

We argue, that the concept of symmetry, and symmetry breaking, can capture fundamental properties of the dynamics of financial markets. At the individual level – the investor – symmetry breaking originates in the cognitive system and can be described as an anchoring. At the social level symmetry breaking is caused by social influence – dependence of others when making investment decisions. At the societal level symmetry breaking is due to feedback loops among different institutional actors who shape financial markets. The idea in the following will then be to argue, that the concept of symmetry can appear in a variety of financial contexts. As such, the hope is it should give new insight to old problems appearing in finance. When the symmetry is preserved, the direction of a market change is unpredictable. Symmetry breaking, in contrast, creates short pockets of predictability, what opens up a chance for arbitrage, i.e., making profit. Specifically, the concept of symmetry/symmetry break will be applied to such examples as technical analysis, long term growth dynamics of financial markets, as well as symmetry of trading strategies applied to buying/selling assets. One of the most interesting aspect of symmetries, is the new understanding one gets, when a symmetry is either broken or restored. The very point where the transition occurs is called a transition of a phase, or just simply a phase transition.

## 2. Symmetry in price formation due to technical analysis

Technical analysis uses past price histories to predict future price movements, whereas fundamental analysis instead uses future estimates of earnings to find the right, fundamental, price at any given time. Below will be described then, how the rise of either unstable support levels, or stable support levels in the price formation introduces symmetries. The understanding of when those symmetries are broken will be shown in certain cases to led to arbitrage possibilities.

### 2.1 Breaking of symmetry from an unstable support level to a stable support level.

Sometimes it can happen the price formation in financial markets can show oscillatory price movements between so-called support levels, for an example, see Figure 2.

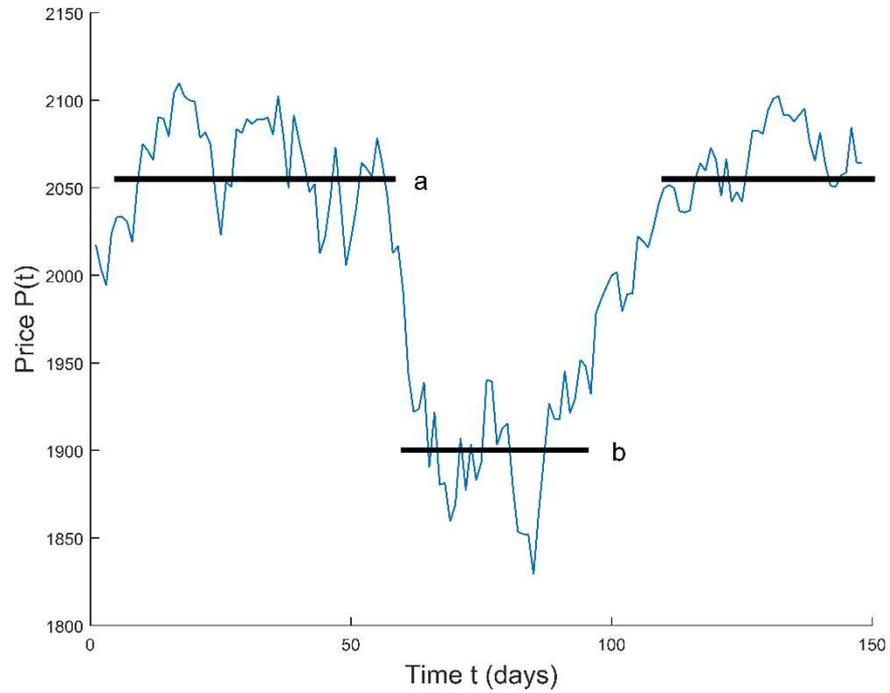

**Figure 2: Symmetry of one stable, and one un-stable support level.** *Illustration of daily price variations, and support levels, of the SP500 index over the period 12/10/2015-13/05/2016;*

The reason for such behavior can be manifold, but one reason could be because of the behavioral trait called "anchoring" [4]. Anchoring describes the tendency for humans to "anchor" on irrelevant information in their decision making. For various works discussion the topic of anchoring see e.g. [5-8]. A simple description taking into account support levels, can be modeled by the equation:

$$\frac{dP(t)}{dt} = \alpha \, [P(t) - a] \, [P(t) - b] \quad (1)$$

Here $a$ and $b$ are the support levels, and $P(t)$ is the price of the asset at time t. $\frac{dP(t)}{dt}$ is the "price velocity" of the asset (for a longer discussion of early warning signals for transitions between support levels, see an excellent article [9]). (1) describes a price velocity that goes to zero when either of the two support levels, $a$ or $b$, are approached, so the two support levels are equilibrium solutions for $P(t) = a$, respectively $P(t) = b$. As can be seen from (1), assuming $\alpha > 0$, $a > b$, then if $P(t) > a$ the price velocity is positive and the prices are "pushed" up and further away from $a$. If on the other hand $P(t) < b$ then the right hand side of (1) is positive, and so prices are "pushed" up towards the support level $b$. For prices in between the two price support levels $a > P(t) > b$, the price velocity is negative so prices are pushed away from $a$ towards $b$.

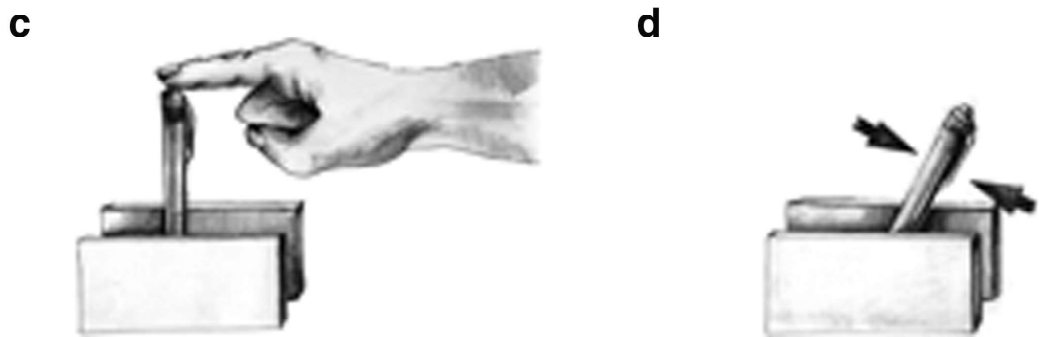

**Figure 3: Two dimensional symmetry break.** *Considering again the pen in Figure 1, but now place two blocks as shown. The pen can then either fall to the left, or fall to the right.*

One can therefore see that, just like the falling pen in Figure 3c, we have a system with an unstable equilibrium $a$, corresponding to the upright pen in Figure 3c. By breaking the symmetry of the unstable solution a, by either a small positive or negative price fluctuation, the system will converge towards a new stable solution $b$ (just like the pen Figure 3d).

On can formally see the appearance of a stable, respectively unstable equilibrium by using linear response theory. Let us call the two equilibria solutions $\overline{P_1} = a$, and $\overline{P_2} = b$. One can then study what happens to a small fluctuation $\varepsilon$ around that price equilibrium. Call $\widetilde{P_1} = \overline{P_1} + \varepsilon$, from (1) we have

$$\frac{d\widetilde{P_1}}{dt} = f(\widetilde{P_1}) \qquad (2)$$

Where the function $f(\widetilde{P_1}) = \alpha\left(\widetilde{P_1} - a\right)(\widetilde{P_1} - b)$ is given by the right-hand side of (1). Linearizing via a Taylor expansion:

$$\frac{d(\overline{P_1} + \varepsilon)}{dt} = f(\overline{P_1} + \varepsilon) \qquad (3)$$

$$\approx f(\overline{P_1}) + \frac{\partial f}{\partial P}\Big|_{P=\overline{P_1}} \varepsilon \qquad (4)$$

Comparing (2)-(4) we therefore get the equation that describes the behavior of the fluctuation $\varepsilon$

$$\frac{d\varepsilon}{dt} = \frac{\partial f}{\partial P}\Big|_{P=\overline{P_1}} \varepsilon \qquad (5)$$

$$\frac{d\varepsilon}{dt} = \mu_1 \varepsilon \qquad (6)$$

With $\mu_1 \equiv \frac{\partial f}{\partial P}\Big|_{P=\overline{P_1}} = \frac{\partial f}{\partial P}\Big|_{P=a} = \alpha\,(a - b)$. Similarly one find that $\mu_2 \equiv \alpha\,(b - a)$. From (6) we then see that for $\alpha > 0$ and $a > b$ any fluctuation around the solution $P(t) = a$ increases exponentially, i.e. the solution is unstable. Similarly any fluctuation around the solution $P(t) = b$ decreases exponentially, i.e. this solution is stable.

It should be noted that the transition from $b$ to $a$, as one can see in Figure 2 cannot be described by (1), so external economic news, or another cause, would be needed in describing such a transition.

2.2 **Symmetry around several stable support levels**

The presence of support levels in financial markets are not limited to equity markets, but can appear in all classes of assets like commodities, currencies and bond markets. One particular case which led to the appearance of support levels, happened in currency markets, and was politically imposed in the 1970's.

After the dissolution of the Bretton Woods system, which had kept the values of currencies fixed to gold, most of the countries of the EEC (the organization prior to the EU) agreed in 1972 to maintain stable exchange rates of their currencies, by allowing for currency fluctuations of at most 2.25%. The fixture was called the European "currency snake". In 1979 the "currency snake" was then replaced by the EMS, the European Monetary System, where most member states linked their currencies in order to prevent large fluctuations relative to one another. The EMS was in place for over a decade, until continued attacks on the system, due to different countries different economic policies, rendered the system obsolete. A much larger fluctuation limit of 15% was subsequently imposed.

However, the fact of imposing relative limits on exchange rates, opens the door for profit taking of traders in the markets. To illustrate how, consider Figure 4 which shows schematically two assets with four different possibilities of price fluctuations around the support levels of asset 1, $\overline{A_1}$, respectively asset 2, $\overline{A_2}$. Inspired by the working of biological motors, L. Gil [10] suggested a trading algorithm to profit from the situation described in Figure 4. In the following the method will be illustrated with only two assets as in Figure 4, but it can easily be generalized to any number of $N$ assets.

**Figure 4: Symmetry of two stable support levels.** *Illustration of a trading algorithm, inspired by how some biologic motors (using a so-called "flashing racket") work by exploiting opportune fluctuations. Each configuration $X_i$ shows one of the 4 different possibilities that can occur for a joint price fluctuations $(dA_1, dA_2)$ around their quasi-static price levels (support levels) $(\overline{A_1}, \overline{A_2})$,.*

The basic idea is for example to go long (i.e. buying) asset 2 whenever a favorable fluctuation happens, as seen for the configuration $X_1$, where a price fluctuation $A_2$ is below the averge price (the support level) $\overline{A_2}$. Next the idea is to wait for another favorable price fluctuation to show up, in this case the configuration $X_4$. When configuration $X_4$ happens, one then sell the asset 2 at a favorable price, and instead go long asset 1 which in configuration $X_4$ is undervalued. This illustrates how the presence of support levels, e.g. politically imposed like the "currency snake", opens up for profit possibilities, by letting a patient trader wait and act when the right price fluctuation happens. However, whether such strategy is actually profitable, and how risky it would be, depends on five things [4]. As shown in [4] for $N = 2$ assets this boils down to:

1) One would need to know the probability to be in any of the 4 configurations $X_i$, $P(X_i)$, shown in Figure 4.

2) One needs to know the transition probability to go from one configuration $X_i$, to another configuration $X_j$, $P(X_i \rightarrow X_j)$.

3) One needs to specify the symmetry break of the algorithm by only seeking out profitable positions. This is for example done by the choice of going long on asset 2 in the case of configuration $X_1$. This is reflected in the probability $P(s|X_i)$ to choose asset $s$ ($= 1,2$) given configuration $X_i$ has occurred. $P(s|X_i)$ is the conditioned probability of choosing asset $s$ given that one has the event of configuration $X_i$.

4) One also needs to know the probability that asset $s$ takes the value $A_i$, given that one is in configuration $X_i$, $P(A_l|X_i)$.

5) Finally one would need to know the transaction costs C.

It is easy to see that in order to have a profitable trading algorithm one should break the symmetry by ensuring to hold assets only in profitable positions. Specifically, this is done by imposing [4]:

$$P(s = 1|X_1) = 0; \quad P(s = 2|X_2) = 1;$$

$$P(s = 1|X_2) = P(s = 2|X_2) = 1/2; \quad (7)$$

$$P(s = 1|X_3) = P(s = 2|X_3) = 1/2;$$

$$P(s = 1|X_4) = 1; \quad P(s = 2|X_4) = 0$$

As shown in [4] the averaged return one would get from such a symmetry break would be:

$$R_{av} = \sum_{i=1}^{4}\sum_{j=1}^{4}\sum_{s=1}^{2} P(X_i)P(X_i \rightarrow X_j)P(s|X_i) \int dA_l \int d A_k \ln\left(\frac{A_l}{A_k}\right) P(A_l|X_i)P(A_k|X_j) \quad (8)$$

The corresponding risk by the trading algorithm can be found by calculating the average standard deviation of the return which is given by:

$$\sigma^2 = E[(r - R_{av})^2] = \sum_{i=1}^{4} \sum_{j=1}^{4} \sum_{s=1}^{2} P(X_i) P(X_i \to X_j) P(s|X_i) \int dA_l \int d A_k \ln\left(\frac{A_l}{A_k}\right)^2 P(A_l|X_i) P(A_k|X_j) - R_{av}^2 \quad (9)$$

After verifying the validity of (8-9), the trading algorithm was then applied to real market situations [4]. However, without political imposed constraints that directly create quasi-static support levels as seen for the currency markets, a general problem arises because of long-term drifts in support levels $\overline{A_1}$ of financial prices. To circumvent this obstacle, the algorithm was modified [4] so as always to be market neutral, independent of any drift the portfolio of the N assets might perform.

To see how this trading algorithm with the induced symmetry break can be applied to real data, consider Figure 5. The figure illustrates how an example of the market neutral algorithm applied to the Dow Jones Industrial Average, as well as the CAC40 stock index. As can be seen from the figure, the algorithm is capable of excellent returns with a high Sharpe ratio. For more information, see [4].

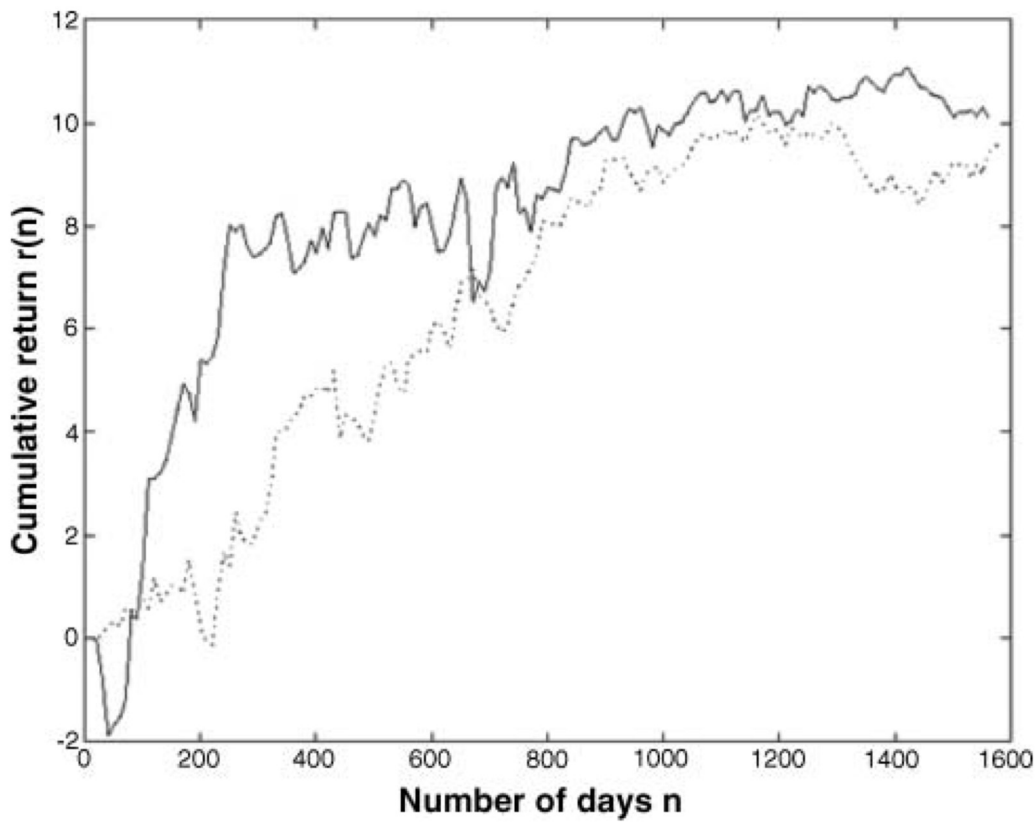

**Figure 5: Trading algorithm applied to a real market situation.** *Cumulative return of the market neutral algorithm applied to daily price data of the Dow Jones stock index (dotted line), as well as the CAC40 stock index (solid line) over the period March 1, 2000–February 5, 2006. First half of the time period for the Dow Jones Industrial Average was used in sample to determine the best choice among three values of the parameter m = 5, 10, 15 days. Second half of the time period for the Dow as well as full period of the CAC40 index was done out of sample with m= 10. As a measure of the performance of the algorithm, the Sharpe ratio was found to be 2.0 and 2.94 for the Dow, respectively, CAC40 price time series. A trading cost of 0.1% was included for each transaction.*

## 3. Symmetry break in trading strategies due to collective decision making

This section will consider, how certain speculative transitions in financial markets can be ascribed to a symmetry break, that happens in the collective decision making. Traditionally economics considers investors to be rational, meaning that in their decision making on whether to buy or sell shares, they use all available information concerning future dividends and future interest rates, to calculate the proper, *fundamental,* price of an asset. A large part of the financial industry therefore has analysts trying to deduce future earnings of companies, recommending to sell if the they deduce the present price is too high, or to buy if the present price is deemed too low when considering future earnings prospective. However, in practice investors use decision making based not only on fundamental price, but use a variety of investment strategies. Technical analysis considers past price performance in order to predict future performance. Investors, and the decision making in their trading of assets, should therefore rather be considered as a dynamically changing pool of investment strategies that depend both on the present price (fundamental value investment strategies) and past price history (technical analysis). In such a setting, markets can be considered as complex systems. As will be shown in the following, the payoff of various investment strategies thereby reflects an intrinsic financial symmetry that generates equilibrium in price dynamics (fundamentalist state), until eventually the symmetry is broken, which then leads to scenarios of bubble or anti-bubble formation (speculative state). If one again considers Figure 3 this illustrates the phenomenological reasoning: the pen standing (i.e. the fully symmetric state) is equivalent to price fluctuations around their fundamental value, whereas the pen falling (the symmetry is broken) is equivalent to price trends towards herding seen in a bubble phase or anti-bubble phase of the market. One can eventually give a behavioral "twist" to such a tale: the financial symmetry breaking corresponds to the financial system moving from one state (or mood), where randomness in price movements obey no-arbitrage conditions, into another state in which the price takes the route towards bubble or anti-bubble formation giving rise to arbitrage possibilities. The first mood is the *fundamental state*, while the second mood is the *speculative state*. A description will be proposed below, that considers such behavior transitions in a micro-to-macro scheme. A market "temperature" will be introduced that modulates the state transitions in the market.

In such a reasoning, an analogy will be made of markets behaving like thermodynamically systems. What characterizes such systems is the struggling forces between energy (order) and entropy (disorder). As will be argued, similar market forces exist forcing the markets to go from one state to another based on their inner "temperature". To make this analogy more clear consider the so-called free energy $F$:

$$F = E - TS \qquad (10)$$

Here $E$ is the internal energy of the system, $T$ the temperature and $S$ the entropy describing how much disorder there is in a given system. From (10) one can see that the state of a system is determined by a struggle between forces representing "order" (the $E$ term), and "disorder" (the $TS$ term).

One note the role of the temperature $T$ in (10) : for $T = 0$ the minimum of the free energy $F$ is just the energy $E$, corresponding to the complete ordered state of the system. However as soon as $T > 0$, the finite temperature will introduce random fluctuations in the system, introducing thereby a non-zero contribution to the entropy $S$. In other words, the temperature describes the amount of randomness in the system. The larger the temperature $T$ the larger this tendency, until at a certain temperature $T_c$ above which order has completely disappeared, and the system is in a disordered state. It is possible to describe such a transition in a macroscopically and phenomenological way, called Ginzburg Landau (GL) theory [11]. In physics, GL theory express the macroscopic properties of a system in terms of a so-called order parameter (see below). As the name suggest, an order parameter can be thought of describing how much order a given system possesses. It could for example be how much spins alignment along a direction for a magnetic material, or for liquid/gas transitions, the order parameter could describe the differences in density of the different phases. It will be suggested to describe in a similar fashion the macro-mechanisms of market transitions, while maintaining consistency in the micro-foundation of individual decision making.

The free energy $F$ in (10) is assumed in GL theory to depend on the temperature T and the magnitude of the order parameter $o$. The assumptions of the GL theory is that the free energy can be expanded in a series expressed in terms of the order parameter $o$:

$$F(T, o) = C + \alpha_2(T)o^2 + \frac{1}{2}\alpha_4(T)o^4 + \cdots \tag{11}$$

It should be noted that (11) does not contain odd terms ($o$, $o^3$,...) in the expansion due to a symmetry argument, for example in physics there is no difference in the free energy for a spin up, respectively spin down system. In finance a similar kind of symmetry exists, directly expressed by the fact that a trader can profit from going long an asset, just as well as going short an asset.

The transition between two phases is triggered by a parameter called the temperature $T$. In order to describe this transition GL assumes:

$$\alpha_2(T) = a(T - T_c); a > 0$$

$$\alpha_4(T) = b = constant > 0. \tag{12}$$

Taking furthermore the derivative in order to find its extreme, we end up with the equation for a minimum of $F(T, o)$, hence determining the state of the system:

$$\frac{\partial F(T,o)}{\partial o} = 2a(T - T_c)o + 2bo^3 = 0, \tag{13}$$

that has the following solutions:

$$2a(T - T_c)o + 2bo^3 = 0$$

$$\text{i)} \quad o(T) = 0 \qquad\qquad T \geq T_c$$

$$\text{ii)} \quad o(T) = \pm\sqrt{\frac{a}{b}(T_c - T)} \qquad T < T_c. \tag{14}$$

It was suggested in [12] to consider the order imbalance, i.e. the difference in volume between buy and sell orders, as an order parameter to characterize the state of a given market. The "free energy" in the financial market context, then corresponds to a "free profit" that traders try to optimize. For more detailed explanations, see [12]. Using such a characterization the solution ii) corresponds to the case of complete order, with all used trading strategies taking the same direction (buy or sell), thereby giving rise to the formation of a bubble or an anti-bubble. This is the pure speculative state, with up or down price movements leading to a bubble or anti-bubble state, breaking the financial symmetry with agents getting positive profits by going long or short in the market. On the other hand, i) corresponds instead to a financial system with disorder. In this case roughly half of the population use strategies that take one action, and the remaining taking the opposite action, thus giving rise to a price path around its fundamental value. This is the disordered state, in which randomness in price movements leads to a no-arbitrage condition.

The *"market temperature"*: One of the main implication of the above mentioned GL-based theory of mood transitions in the market, is the existence of a nontrivial transition from a "high temperature" symmetric state, where traders don't create a trend over time, to a "low temperature" state, characterized by trend following with a definite trend in the price trajectories (up or down). In the following a "temperature" linked to the randomness of the agent's actions will be suggested via agent based modeling. In [13] an agent based model was used in which the agents used fundamental analysis strategies, as well as technical analysis strategies. Randomness entered the modeling via initial conditions in the assignments of $s$ technical analysis strategies to $N$ traders in the game. The technical analysis strategies were created by randomly assigning either a -1 (meaning sell) or a 1 (meaning buy) for each of the $2^m$ different price histories, that one gets from considering only the last m up (1) or down (0) price movements. In such a modeling the total pool of strategies increases as $2^{2^m}$ versus $m$. However, many of these strategies are closely related, and refs. [14-15] showed how to construct a small subset of size $2^m$ of independent strategies out of the total pool of $2^{2^m}$ strategies. As suggested in Refs. [16-18], a qualitative understanding of this problem can be obtained by considering the parameter $\alpha = \frac{2^m}{N}$. Along the same line of reasoning, Ref. [19] pointed out, however, that the ratio $\alpha = \frac{2^m}{N \times s}$ seems more intuitive, since this quantity describes the ratio of the total number of relevant strategies to the total number of strategies held by the traders. In similar vein a "market temperature" was introduced in [13] via the expression:

$$T = \frac{2^m + 1}{N \times s}, \tag{15}$$

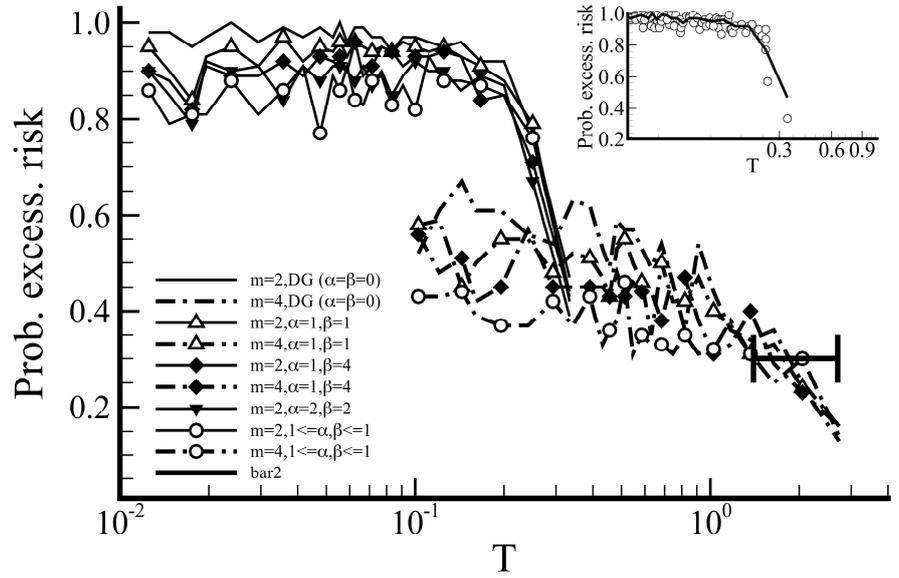

where the "+1" in the numerator is because of the fundamental strategy in addition to the $2^m$ uncorrelated speculative strategies.

From (15) one can then see how the mechanism behind a "temperature" that can break the symmetry and shift the market between different moods/states: when the total pool of strategies held by the $N$ traders is small with respect to the total pool of relevant strategies (low denominator of $T$), this corresponds indeed to the large fluctuations, large temperature case. Vice versa a large pool of strategies held by the $N$ traders (increasing denominator of $T$), therefore corresponds to a small temperature case.

**Figure 6: The importance of T: probability of excessive risk-taking versus T for groups with different risk profiles.** Collective speculative price behavior, defined as occurring when the final price is more than twice the fundamental price, is plotted against the control parameter $\frac{[3^M+1]}{(s+1)N}$. Inset: simulations of games with different values of s and N, but for a fixed value of T and M≡2. The different symbols correspond to simulations with populations having different risk profiles.

Figure 6 shows Monte Carlo simulations of an agent-based model [13] illustrating the relevance of the "temperature" introduced above. On the y-axis is plotted the probability of generating a collective speculative price behavior against the control parameter $T$, different curves correspond to different risk profiles of the agents. The first thing to notice is by considering the inset which shows that given *fixed* values of M and T lead to the *same* probability of creating excessive price behavior (see Figure 6 inset, which presents simulations of games with *different* values of s and N, but for *fixed* T and M≡2).

The simulations shown in Figure 6 were carried out for agents using their total return as payoff function (the $-Game [20]), but with different risk profiles (different curves). From Figure 6 one can see that different risk profiles gave similar behavior, but the temperature defined above could indeed switch the micro behavior of agent from speculative behavior (small T) to fundamental behavior (large T). The two different clusters of curves correspond to two different sizes in the price histories used by the agents (see [13]).

In practice one can use the idea of a symmetry break try to detect the transition from fundamental to speculative behavior in financial market experiments (for some literature of related, but more traditional experiments see also [21-25]). Figure 7 show the price history versus time (topmost plot) for one out of ten experiments, as well as (bottom plot) a so-called "decoupling" parameter (defined via a subset of strategies, for a detailed description see [26]) versus time obtained via Monte Carlo (MC) simulations. The MC simulations were created using real experimental price formation data, by "slaving" the price history from the experiments as input to agents in the $G simulations. That is, instead of having the agents reacting to their own repeated-game actions, the agents in the simulations would instead use the actions (price history) of the participants. The important point to notice, is that by looking at the decoupled strategies (bottom plot of Figure 7), the split in the use (the percentage) of the two types of strategies heralds the onset of the breaking of a symmetry, which as seen from the top plot, is confirmed by the speculative price formation made by the participants in the experiment.

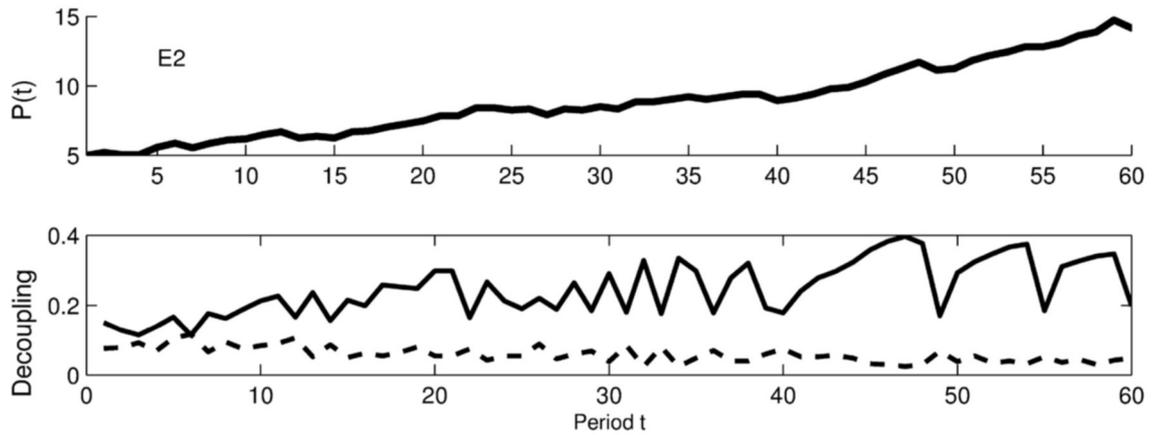

**Figure 7: Breaking of symmetry by decoupling of strategies.** *Cumulative Price formation in experiments (top plot) and decoupling of strategies in agent based modeling "slaved" to the experimental price (bottom plot). The split seen in the decoupled strategies marks the onset of speculation seen in the experiment. For a detailed description see [26].*

4. **Symmetry and its role for the long term growth of financial markets**

In the following the concept of symmetry will be studied in a different setting, namely the phenomenon of long term growth of financial markets. It will be argued that in this case, the human decision making on whether to go long (buy a share) or go short (sell a share) introduces a symmetry relevant for long term behavior of markets. The question posed is: what conditions are needed for a long term sustainable growth or contraction in a financial market? It will then be argued, that the concept of symmetry is relevant in the answer to this question, because of an inherent asymmetry between the role of traditional market players of long only mutual funds, versus the role of hedge funds, who take both short and long positions. If one considers the traditional composure of market participants, then by far a majority of participants (pension funds, insurance companies, retail investors, ….) are only allowed to take long (buy) positions. However, since the beginning of the 1990 a steadily increasing percentage of market participants are hedge funds. They do not have the same limitations as traditional funds, and can take up long as well as short positions. The question that arises is then, as to what effect this "broken" symmetry of a historically surplus of long-only market participants could mean for the long term behavior of financial markets as well as the economy?

### 4.1 The asymmetric case: one large dominating fund (a mean-field solution)

Let us therefore, consider the case of a given financial market, which for simplicity will be assumed to be controlled by just one large investor/fund that tries to maintain a certain constant growth rate of the market. If one call the wealth of the fund at time t, for $W(t)$:

$$W(t) = n(t)P(t) + C(t) \quad (16)$$

$n(t), P(t)$ are the number of shares held by the fund and the price of the shares at time t, respectively. $C(t)$ is the cash possessed at time t by the fund. Since the aim of the fund is to ensure a constant growth rate $\alpha$ of the market, the fund at every time step needs to keep on buying a certain number of shares $n(t)$. In order to see how many shares $n(t)$, are needed at time t, let us call the excess demand of shares created by the constant buying for $A(t)$. In the following it will be taken a constant in time, $A(t) \equiv A$. Therefore:

$$\frac{d\,n(t)}{dt} = A; \quad n(t) = A\,t \quad (17)$$

Since the return of the price, $R(t)$, is proportional to the excess demand one has:

$$R(t) = \ln\left(\frac{P(t+1)}{P(t)}\right) = \frac{A(t)}{\lambda} \; ; \quad P(t) = e^{\frac{At}{\lambda}} \quad (18)$$

With $\lambda$ the liquidity of the market.

Apart from the expenses of the fund to keep on buying shares, the fund get an income from the interest rate $r(t)$ on its cash supply $C(t)$, as well as an income from the dividends of the shares $d(t)$. The balance equation for the cash supply $C(t)$ reads:

$$\frac{d\,C(t)}{dt} = -\frac{dn(t)}{dt}P(t) + C(t)r(t) + n(t)d(t) + C_{flow}(t, r(t), d(t), P(t), \dots) \quad (19)$$

Where $C_{flow}(t, r(t), d(t), P(t), ...)$ takes into account the additional cash inflow/outflow which, as indicated, could for example depend on the time, the return, the dividends or the price of the market. Since the idea is to study the ability of the fund to maintain a certain growth rate $\alpha \equiv \frac{A}{\lambda}$, it makes sense to rewrite (19) in the following way:

$$\frac{d\tilde{C}(t)}{dt} = -\alpha e^{\alpha t} + \tilde{C}(t)r(t) + \alpha t d(t) \quad (20)$$

In (20) the cash is now renormalized in terms of the market liquidity $\tilde{C} \equiv \frac{C}{\lambda}$, the interest rate is assumed constant, $r(t) \equiv r$, , and for simplicity the cash flow term, $C_{flow}$, has been left out in (20).

In [27] it was shown that for the case of constant dividends, $d(t) \equiv d$, the fund could not maintain a "super-interest" growth, $\alpha > r$, but in certain cases it was found that the fund could enable "sub-interest" growth, , $\alpha < r$. For further elaboration on the topic see [27]. In order to obtain a "super-interest" growth, as seen in the booming 1990s and more recently, one would need to include the impact of the so-called wealth effect. In (20) this is taken into account by by assuming a positive feedback of the stock price on the earnings of a firm, so that the dividend is assumed to increase proportional to the price $d(t) = \frac{d_0 P(t)}{P(t=0)}$. Using this assumption (20) takes the form:

$$\frac{d\tilde{C}(t)}{dt} = -\alpha e^{\alpha t} + \tilde{C}(t)r(t) + \alpha t e^{\alpha t} d_0 \quad (21)$$

With the solution

$$\tilde{C}(t) = \alpha e^{\alpha t} \left\{ \frac{td_0 - 1}{\alpha - r} - \frac{d_0}{(\alpha - r)^2} \right\} + e^{\alpha t} \left\{ \frac{-r\alpha + \alpha^2 + \alpha d_0}{(\alpha - r)^2} + \tilde{C}(t = 0) \right\} \quad (22)$$

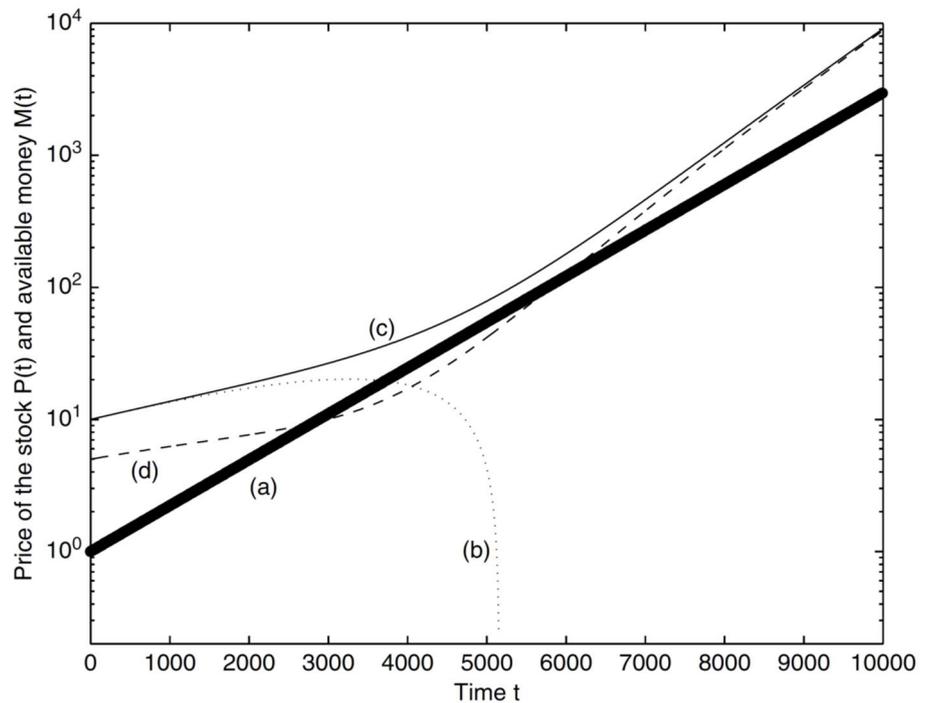

**Figure 8: When is long term growth of markets possible?** *a) Price P(t) assuming constant growth rate. b) Cash solution of (21) which isn't able to sustain the growth of curve a) (the price of the market at some point in time becomes larger than the cash available). c) sufficient initial amounts of cash can however enable constant growth. For a detailed description see [27].*

In order to see how a positive feedback of the stock price on the earnings of a firm, could make super-interest growth possible consider first curve b) in Figure 8. The curve illustrates a solution of (21) where a given amount of initial cash, initial dividends and an interest rate r of 10%, is *not* enough to sustain a super-interest growth of the stock market, since as can be seen, the fund runs out of cash. If on the other hand one considers curve c), which is for market behavior with the same conditions as b) except having a higher initial dividend d0 of 8%, then a super interest growth of 20% of the markets *does* become sustainable (this is seen since the cash as a function of time is always larger than the price of the market, $C(t) \geq P(t) \forall t$). Sufficient initial amount of cash is however required, as illustrated by curve d), where a high initial dividend is not sufficient to avoid that the investor runs out of money (C(t) < P(t) for t ≈ 2800).

The exact same exercise can be done for the other asymmetric case of a fund that instead tries to profit from keeping pushing the market down by a steady rate of short selling of shares, leading to a negative growth $-\alpha$ of the market. For a longer discussion of this case, see [27]. To summarize: the solution of (21) illustrates how a broken symmetry created by having an excess of investors trying either to push up/push down the market is not only possible, but also profitable for such investors. Considering price evolution of financial markets as a long term growth phenomenon, (21) illustrates how a broken symmetry of market participants taking up more long, or more short positions, could lead to a long term growth/decline of the markets. As argued above, financial markets have always been in the situation of having a broken symmetry tilted towards buying market participants, due to the mere investment strategies of a majority of market participants. The question then is how a restoration of this broken symmetry could influence the market price formation?

### 4.2 The symmetric case: equal amounts of market participants buying/selling shares (agent based simulations)

Equations (19)-(22) express the case of one single powerful investor that has enough liquidity to drive the market. This is clearly not a very realistic situation. Consider instead the much more realistic and complex case of a group of investors, that all try to drive the market for their own profit. In order to see, if a group of agents with heterogeneous trading strategies would change the conclusion of Section 4.1, we then consider results from agent based simulations, Figure 9. The figure illustrates one price history (fat solid line) for simulations of an ensemble of agents in the so-called $-game [20]. At the beginning of each game, a fraction $\rho$ of the agents were allowed to take both long and short positions; the remaining $1-\rho$ fraction of agents were allowed only to take long positions. Thin dotted lines represent the 5%, 50% and 95% quantiles (from bottom to top) respectively, i.e., at every time *t* out of the 1,000 different initial configurations only 50 got below the 5% quantile line, 500 got below the 50% quantile and 950 below the 95% quantile. All the agents in Figure 9a were long-only and so correspond to the analysis

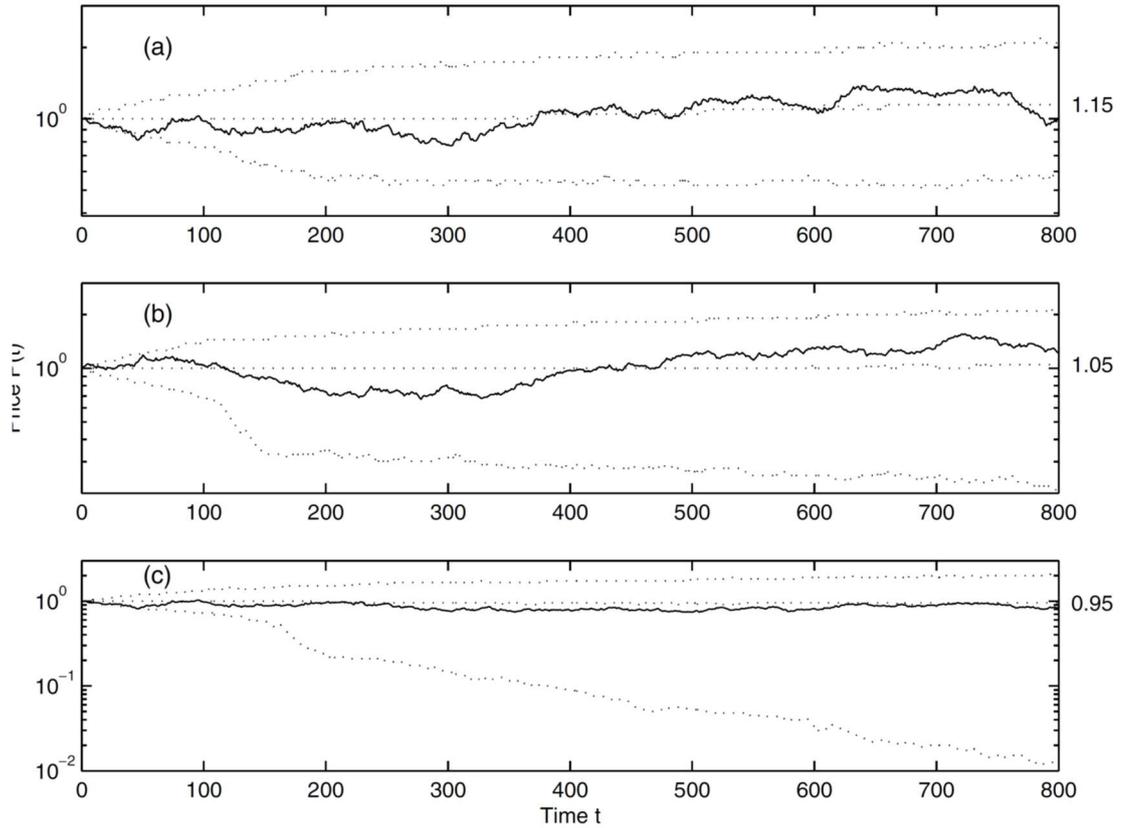

performed in section 4.1. As can be seen from Figure 9a the *average* behavior of an agent can now be understood using the analysis of equations (19)–(22) since the price $P(t)$ is *in average* tilted towards positive returns. Figure 9b represents simulations with the same parameter values as in Figure. 9a except that a fraction, $\rho = 0.20$ of the agents can be both short and long. As can be seen from the 5% quantile, the introduction of agents that can take short positions clearly increases the probability significantly for a lasting bearish trend. Increasing $\rho$ to 0.4 in Figure 9c amplifies this tendency

**Figure 9: Restoring the symmetry via agent based simulations.** *Fat solid line: price P(t) from one configuration of the $-game [27]. Thin dotted lines represent the 5%, 50% and 95% quantiles (from bottom to top) respectively, i.e., at every time t out of the 1,000 different initial configurations only 50 got below the 5% quantile line and similarly for the other quantile lines. (a) the fraction of agents allowed to take short positions, ρ = 0, (b) ρ = 0.2 and (c) ρ = 0.4. For a detailed description of the simulations see [27]*

## 5. Concluding remarks

In this article we have suggested to put forward using the idea of symmetry as a tool to apply in a financial context. The concept of symmetry has had a huge impact in the way physicist think about problems in most sub branches of physics, and presenting this article, it is our hope the concept could be of similar use for practitioners/theoreticians working in the field of finance. We have argued for the importance of the symmetry of strategies that can happen in the decision making to buy or sell financial shares. As shown game theory therefore is one obvious avenue where to look for symmetry. It was shown how the classical order vs. disorder phase transition problem in physics, could be used to explain the fundamentalist vs. speculative mood transitions in the markets, something that was proposed to disentangle via a Ginzburg-Landau-based power expansion. In this case the key parameter that breaks the symmetry and moves the markets from one state to another is the "temperature parameter", which we derive based on the randomness of the agent's actions, the number and the memory length of traders. The financial symmetry breaking in this case corresponds to the financial system moving from one state (or mood), where randomness in price movements obey no-arbitrage conditions, into another state in which the price takes the route towards bubble or anti-bubble formation giving rise to arbitrage possibilities [28]. We have demonstrated how symmetry breaking occurs at the individual level though anchoring, and the collective level though collective decision making. Finally, we also showed how long term economic growth could be understood as a phenomenon determined the concept of symmetry.

Acknowledgments

AN contribution was supported by funds from Polish National Science Centre (project no. DEC-2011/02/A/HS6/00231).